# Topologically ordered magnesium-biopolymer hybrid composite structures

Reece N. Oosterbeek[a,*], Christopher K. Seal[b], Mark P. Staiger[c], Margaret M. Hyland[a]

[a]Department of Chemical and Materials Engineering, The University of Auckland, Private Bag 92019, New Zealand

[b]Light Metals Research Centre, The University of Auckland, Private Bag 92019, New Zealand

[c]Department of Mechanical Engineering, University of Canterbury, Private Bag 4800, Christchurch 8020, New Zealand

Abstract

Magnesium and its alloys are intriguing as possible biodegradable biomaterials due to their unique combination of biodegradability and high specific mechanical properties. However, uncontrolled biodegradation of magnesium during implantation remains a major challenge in spite of the use of alloying and protective coatings. In this study, a hybrid composite structure of magnesium metal and a biopolymer was fabricated as an alternative approach to controlling the corrosion rate of magnesium. A multistep process that combines metal foam production and injection moulding was developed to create a hybrid composite structure that is topologically-ordered in all 3 dimensions. Preliminary investigations of the mechanical properties and corrosion behaviour exhibited by the hybrid Mg-polymer composite structures suggest a new potential approach to the development of Mg-based biomedical devices.

1.  Introduction

There has been strong recent interest in magnesium (Mg) and its alloys as potential orthopaedic implant materials due to its unique characteristic of biodegradability *in vivo*. The biocompatibility of Mg ions within the body has been indicated [1-3], along with some evidence that Mg may be osteogenic [4-8]. In addition, the elastic modulus of Mg closely resembles that of bone which is a requisite for minimising the stress shielding of bone tissue that surrounds an implant [9]. However, a challenge in developing biodegradable materials including Mg is in achieving uniform and controlled degradation of the implant over its lifetime [9-11].

Extensive research has been carried out recently to investigate the use of coatings for controlling the corrosion rate of Mg *in vivo* [11, 12]. However, the various coatings trialled on Mg substrates present their own specific challenges that may result in a combination of mismatched of mechanical properties, poor adhesion to the substrate, exacerbation of localised corrosion, non-uniform degradation, and/or creation of debris particles. Attempts to allow slow corrosion rate of Mg by use of porous or permeable coatings is problematic [13-15]. In particular, the release of $H_2$ gas during corrosion results in blistering and delamination of the coatings [16-19].

In this work, a novel approach is investigated as an alternative means of controlling the corrosion rate of Mg. It is hypothesised that an interpenetrating network composite, consisting of a magnesium scaffold and a suitable degradable polymer, can give uniform dissolution from outside-in. This would retain mechanical strength, giving a controlled corrosion rate and the avoiding accelerated corrosion that occurs with coatings when the coating is breached [11, 16]). This work demonstrates the production of such a composite via injection moulding of a porous magnesium scaffold produced by rapid prototyping and casting. The mechanical properties and corrosion behaviour of this composite are then examined.

2.  Experimental procedures

*2.1  Preparation of topologically-ordered porous magnesium*

Firstly, porous Mg with an ordered topology was fabricated as previously described elsewhere [20, 21]. Briefly, a 3D computer model of the desired porous structure is generated (Figure 1a). The ordered structure is then fabricated using a polymer *via* 3D printing to create a positive template. The polymer template is infiltrated with a NaCl paste. Subsequent heating of the template burns off the polymer while also sintering the NaCl. The negative NaCl template is then pressure-infiltrated with liquid Mg (99.8%, Progressive Casting, NZ). Finally, the NaCl template is removed by washing in $Na_2SO_4$ solution. Cylindrical porous Mg samples (20 mm long, 20 mm diameter) were fabricated having 2 mm square pores and 1.5 mm thick struts to give a theoretical porosity of 59.6% (see Figure 1b).

*2.2  Fabrication of magnesium-biopolymer composite*

A Mg-biopolymer hybrid structure was fabricated by injection moulding the porous Mg structure described above with a polylactide (PLA 3051D, NatureWorks® LLC). The polymer feedstock was dried at 70 °C for 20 h prior to use to prevent the hydrolysis of PLA during injection moulding [22]. The porous Mg scaffolds were placed inside a steel mould, with a cylindrical cavity (diameter 20mm, height 20mm) and injection moulded with PLA (Mini-Jector Injection Molder, Mini-Jector Machinery Corp., USA). The cylinder and nozzle temperatures were 212 °C, and 208 °C, respectively. Injection was carried out at a pressure of 1200 psi with an injection time of 12 s.

*2.3  Corrosion analysis*

Immersion tests were performed to semi-quantitatively assess the differences in corrosion rate between pure Mg (99.8%, Progressive Casting, NZ) and the Mg-PLA composites. Test samples were produced as discs with a diameter of 18 mm and length of 8 mm. The surfaces of all samples were ground with P1200 SiC paper prior to testing. The pure monolithic Mg samples were prepared with a surface area (~9.6 cm$^2$) equivalent to the Mg-PLA composites. Samples were immersed in 80 mL of 3 wt.% NaCl solution for up to 2 weeks at 37 °C (±0.5). The surface area to fluid volume ratio was kept constant at 0.12 cm$^{-1}$ for all testing. The pH of the solution was not buffered during testing but was monitored approximately every 15 hours using a digital handheld pH meter (which was calibrated before each use using standard solutions with known pH). One of the authors (MPS) has shown that various buffers, commonly used to control pH during the biocorrosion of Mg, will themselves influence the corrosion behaviour of Mg [23]. Thus, in this work buffers were not considered since only a relative comparison between pure Mg and Mg-PLA composites was required. The samples were removed from the solution after 2 weeks and rinsed with distilled water and ethanol, and dried with hot air.

Additionally, an estimate of the corrosion behaviour was obtained by measuring the % reduction in area of the corroded samples. Macro-photographs were taken of the corroded samples, calibrated to known distances, and the % reduction in area calculated using the image analysis software ImageJ.

2.4  *Mechanical testing*

The compressive 0.2% offset yield strengths of pure Mg and Mg-PLA hybrid composite structures were measured based on 3 replicates. A uniaxial testing machine was used (Instron 5567) equipped with a 30 kN load cell. Strain was measured from the crosshead displacement and the crosshead speed was 0.018 mm/s. Samples were prepared with a diameter (D) of approximately 15 mm and length (L) of 18 mm, giving an L/D ratio of 1.2 in line with ASTM Standard E9-09 [24]. The friction between sample and platens was minimised by grinding sample surfaces with SiC paper (P1200) and using a lubricant (MoS$_2$, Molykote®, Dow Corning Corporation).

*2.5    Materials characterisation*

Mg-PLA samples were mounted in epoxy resin under vacuum in order to preserve the microstructural features of the corrosion layer. Samples were then polished to a 1μm finish. Accelerating voltage varied between 5-10 kV, and a solid state detector (for backscattered electrons) was used. All samples were sputter-coated with a 10nm layer of platinum prior to electron microscopy. Scanning electron microscopy (SEM) and energy dispersive spectroscopy (EDS) were carried out with an FEI Quanta 200 FESEM.

3.    Results and discussion

*3.1    Hybrid composite structures based on Mg and PLA*

The porous Mg structure was completely infiltrated by PLA during injection moulding. Voids in the PLA or at the Mg-PLA interface could not be observed in any of the cross-sections by SEM examination. The measured density of the composite (1.44 ± 0.04 g/cm$^3$) was very close the theoretical value (1.43 g/cm$^3$), confirming near full infiltration of the porous Mg structure.

The rough surface of the Mg shown in SEM images is a result of both the casting method used, and corrosion of the Mg surface - this corrosion was caused during the washing out of the NaCl template using Na$_2$SO$_4$. The resulting corrosion products were removed by washing with chromic acid, leaving a rough surface. EDS analysis of the Mg surface after washing with chromic acid showed no residual chromium or other effects aside from removal of corrosion products. To ensure the structures were free of micro-voids, cross sections were examined at three different points within each sample that corresponded distances of 0, 10 and 20 mm from the injection point. A small number of microvoids (approximately 5 μm) were observed at the Mg-PLA interface furthest from the injection point (20 mm), indicating that the injection moulding process requires further optimisation to produce completely defect-free structures (Figure 2).

*3.2    Compressive behaviour of Mg-PLA hybrid composite structures*

The compressive yield strengths of the Mg-PLA composite and pure monolithic Mg were 71 ±2 MPa and 113 ±10 MPa, respectively. The incorporation of PLA into the hybrid structure clearly decreases the mechanical strength of the Mg due to its lower yield strength (*e.g.* tensile strength 48.3 MPa [25]) compared with Mg. The yield strength of the Mg-PLA composite measured here closely matches the yield strength that can be calculated based on a simple rule of mixtures, taking into account the yield strength of the two component materials – 74MPa. The compressive yield strength of pure Mg is similar to that seen in the literature [9].

Stress strain curves for the bulk Mg and Mg-PLA composite are shown in Figure 3. It is interesting to note that the stress-strain curve for the Mg-PLA composite is similar to characteristic curves for metal foams [26, 27], with an initial elastic region, followed by stress plateauing after the material yields, before a further increase due to the cross-sectional area increasing under compression.

*3.3    Corrosion behaviour of Mg-PLA hybrid composite structures*

Corrosion of the Mg-PLA hybrid composite structure appears to be relatively coherent, from the outside inwards (Figure 4), rather than fluid being channelled throughout the composite structure along the Mg-PLA interfaces via capillary action (where the surface tension of the fluid draws it up along the interface between the Mg and PLA), which would cause corrosion across the whole sample. This is an important finding as it demonstrates that the adherence of the PLA to the Mg substrate prevents the passage of the corrosion medium into the interior of the hybrid structure. In comparison, many of the current coating technologies will prevent corrosion for a period before failing and then exposing the bare Mg substrate to the corrosion medium [16].

The corrosion products (marked "C" in Figure 5) were determined be $Mg(OH)_2$, as only Mg and O were detected in the EDS scans, and $Mg(OH)_2$ is known to be the typical corrosion product of Mg [12]. Figures 5a and 5b show a PLA strut surrounded by Mg; it can be seen that a small amount of corrosion has occurred around the edge of the PLA. This area was located in the centre of the sample and was therefore not in direct contact with the corrosion solution, so exposure and subsequent corrosion is thought to have occurred by the PLA absorbing corrosion solution (a known phenomenon [28, 29]).

Contact with a limited amount of corrosion solution via either of these mechanisms explains the small corrosion layer: as the Mg corrodes, it forms $Mg(OH)_2$, resulting in a volume expansion (as $Mg(OH)_2$ has a higher specific volume than Mg - 24.3 $cm^3$/mol compared to 14.0 $cm^3$/mol [30]). This volume expansion (of 0.42$cm^3$/g(Mg)) results in the PLA pressing the corrosion layer back against the Mg, helping it remain intact to form a protective layer. This observation can also explain results seen above in Figure 4, where corrosion was seen to take place from the outside in, rather than along the Mg-PLA interfaces - the volume expansion may create a seal, preventing corrosion fluid from travelling along the interfaces (in addition to preventing capillary action).

Figures 5.c-e show examples of corrosion occurring from the edge of the Mg-PLA composite sample inwards, as observed in macro photographs (Figure 4). It can be seen that the Mg component of the composite samples exhibits a similar corrosion mechanism as bulk Mg (shown in Figure 5.f), namely localised pitting corrosion in some areas, and minimal general corrosion in other areas. There does not appear to be any evidence of corrosion attack being accelerated in areas with close proximity to the PLA components.

The corrosion test pH trace is shown in Figure 6; each line is an average of at least three measurements. Two major features are demonstrated in this graph. Firstly, both materials show a gradual decreasing trend, and secondly, the pH of the Mg-PLA composite is consistently lower than that of the bulk Mg.

The gradual decreasing trend is thought to be a result of dissolved corrosion products affecting equilibrium reaction rate kinetics (see corrosion reaction equations 1 and 2). Dissolved $MgCl_2$ accumulates, slowing down the conversion of $Mg(OH)_2$ to $MgCl_2$, which in turn slows the corrosion of Mg into $Mg(OH)_2$. This is probably also the cause of the sharp increase in pH in the early stages of the corrosion test (<10 hours) - there are no corrosion products in the solution to begin with, so the reactions can progress rapidly, until they accumulate and slow the forward reactions down.

$$Mg + 2H_2O \leftrightarrow Mg(OH)_2 + H_2 \qquad (1)$$

$$Mg(OH)_2 + 2Cl^- \leftrightarrow MgCl_2 + 2OH^- \qquad (2)$$

The other main feature is that the pH of the Mg-PLA composite is consistently lower than the bulk Mg. This would usually indicate reduced corrosion however, dimensional changes indicated increased corrosion for the Mg-PLA composite (see below). Therefore it is more likely that the reduced pH is due to degradation of the PLA into lactic acid. This is the normal breakdown route of PLA and is well documented in literature [28, 29]. This pH reduction is thought to be the cause of the increased corrosion experienced by the Mg-PLA composite, as Mg corrosion is known to be accelerated at lower pH values [31, 32]. No dramatic changes in behaviour were observed over the course of the test, indicating that both the Mg-PLA composite and pure Mg undergo constant corrosion.

The % reduction in area for the pure Mg and the Mg-PLA composite were also used to compare the amount of corrosion taken place, and are shown in Figure 7. It is clear that the Mg-PLA composite has undergone significantly more corrosion than the pure Mg sample; this is due to the degradation of PLA into lactic acid, reducing pH and therefore accelerating Mg corrosion [28, 31, 32]. Although this study has not been successful in reducing the corrosion rate of Mg, it has shown that this type of Mg-polymer interpenetrating network composite is resistant to bulk corrosion and instead corrodes from the outside in. This paves the way for future developments that could make use of alternative polymers to alter degradation behaviour.

4.  Conclusions

Method for the production of Mg-polylactic acid interpenetrating network composites was developed, building on an existing method for the production of porous Mg and using injection moulding for the infiltration of PLA.

Mechanical tests (specifically compression tests) were carried out on the Mg-PLA composite and compared with bulk Mg. It was found that the composite demonstrated a lower yield strength of 71 ±2 MPa, compared to 113 ±10 MPa for bulk Mg. The bulk Mg was seen to exhibit characteristic shear failure, while the Mg-PLA composite failed by crushing and collapse of the Mg struts.

One promising result seen was the corrosion mechanism of the Mg-PLA composite. It was seen to corrode from the outside inwards, rather than the corrosion fluid permeating the structure (along the Mg-PLA interfaces or via absorption into the PLA) and causing bulk corrosion. This is thought to be a result of the volume expansion that occurs when Mg reacts to form $Mg(OH)_2$, which in turn causes the PLA to press back on the $Mg(OH)_2$ layer, forming a seal and keeping the corrosion layer intact and protecting the underlying Mg from further corrosion.

Corrosion tests were carried out (by immersion in 3% NaCl at 37°C for 2 weeks). These showed greater corrosion of the Mg-PLA composite than bulk Mg, which is attributed to breakdown of PLA into lactic acid, lowering pH (confirmed by pH monitoring) and thus accelerating Mg corrosion. This suggests that PLA may not be the most suitable polymer for use in this composite system.

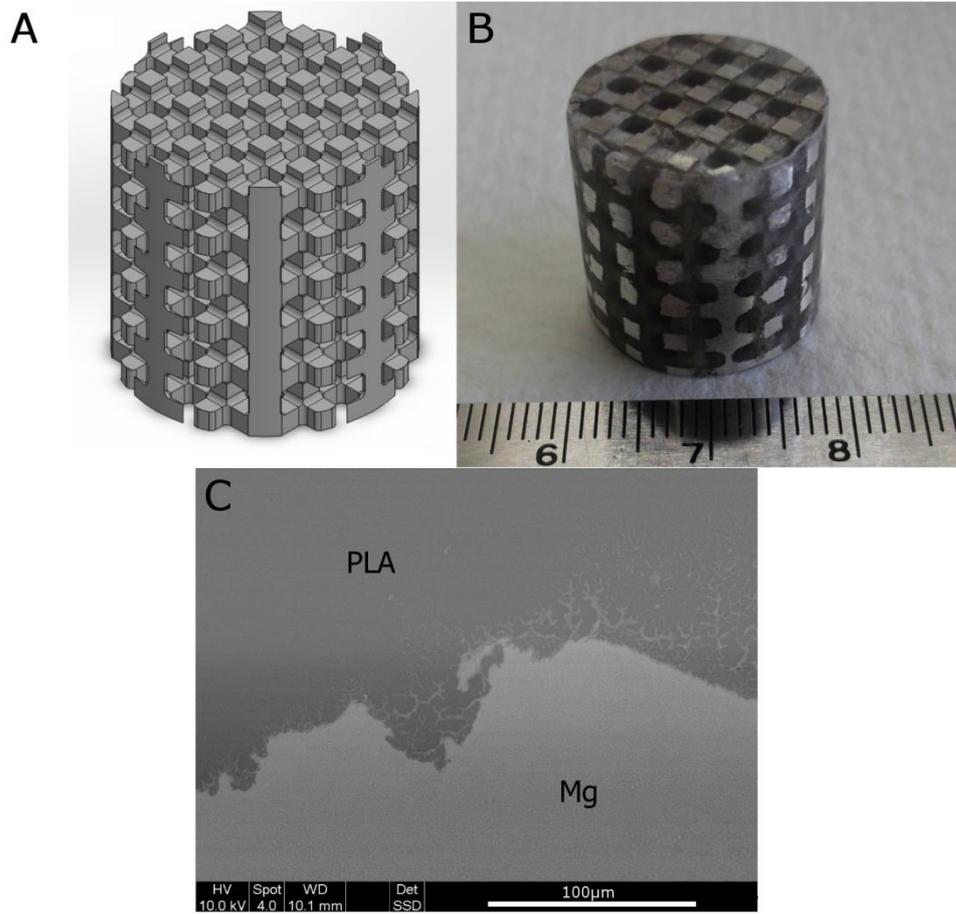

Figure 1: (a) CAD model of the porous Mg scaffold, (b) photograph of the final porous Mg structure, and (c-d) scanning electron micrographs of the cross-section of the final Mg-PLA interpenetrating composite structure.

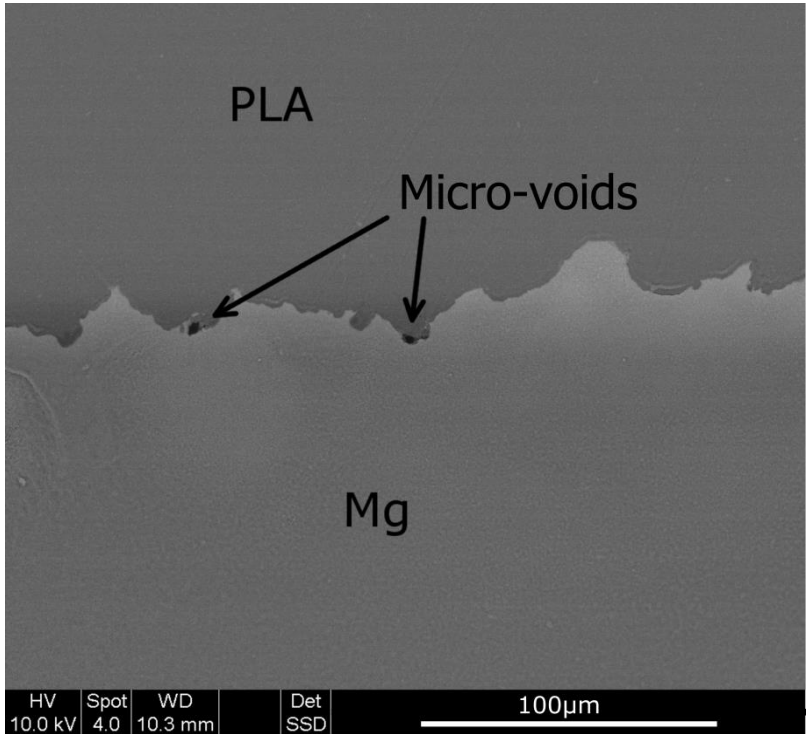

Figure 2: SEM image of the Mg-PLA composite, showing micro-voids present at the interface between Mg and PLA, at maximum distance of 20 mm from the injection point.

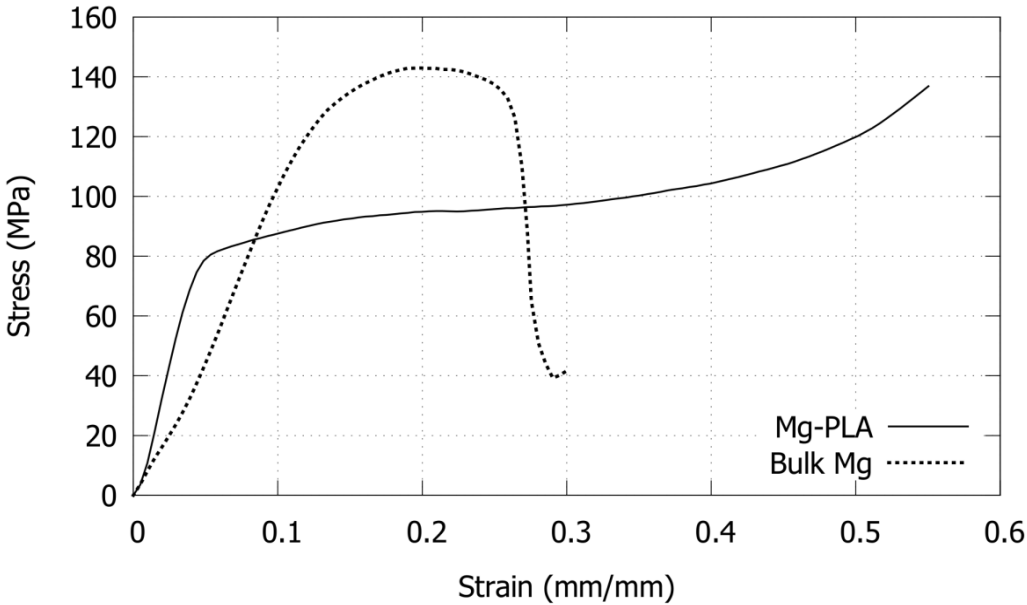

Figure 3: Stress-strain curves for the Mg-PLA composite (solid line) and bulk Mg (dotted line).

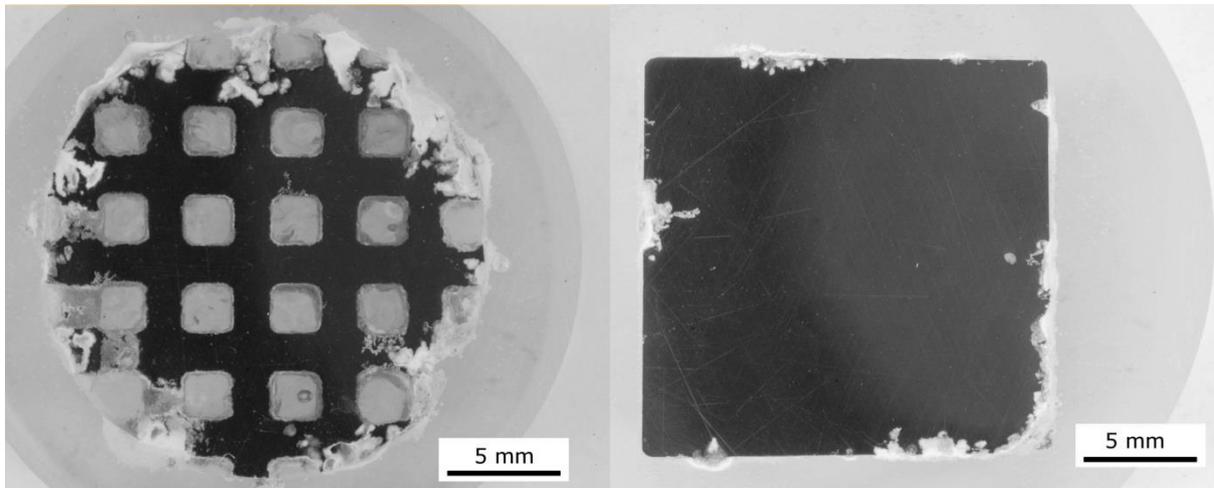

Figure 4: Cross-sectional view of the corroded Mg-PLA composite (left) and bulk Mg (right) after the immersion test.

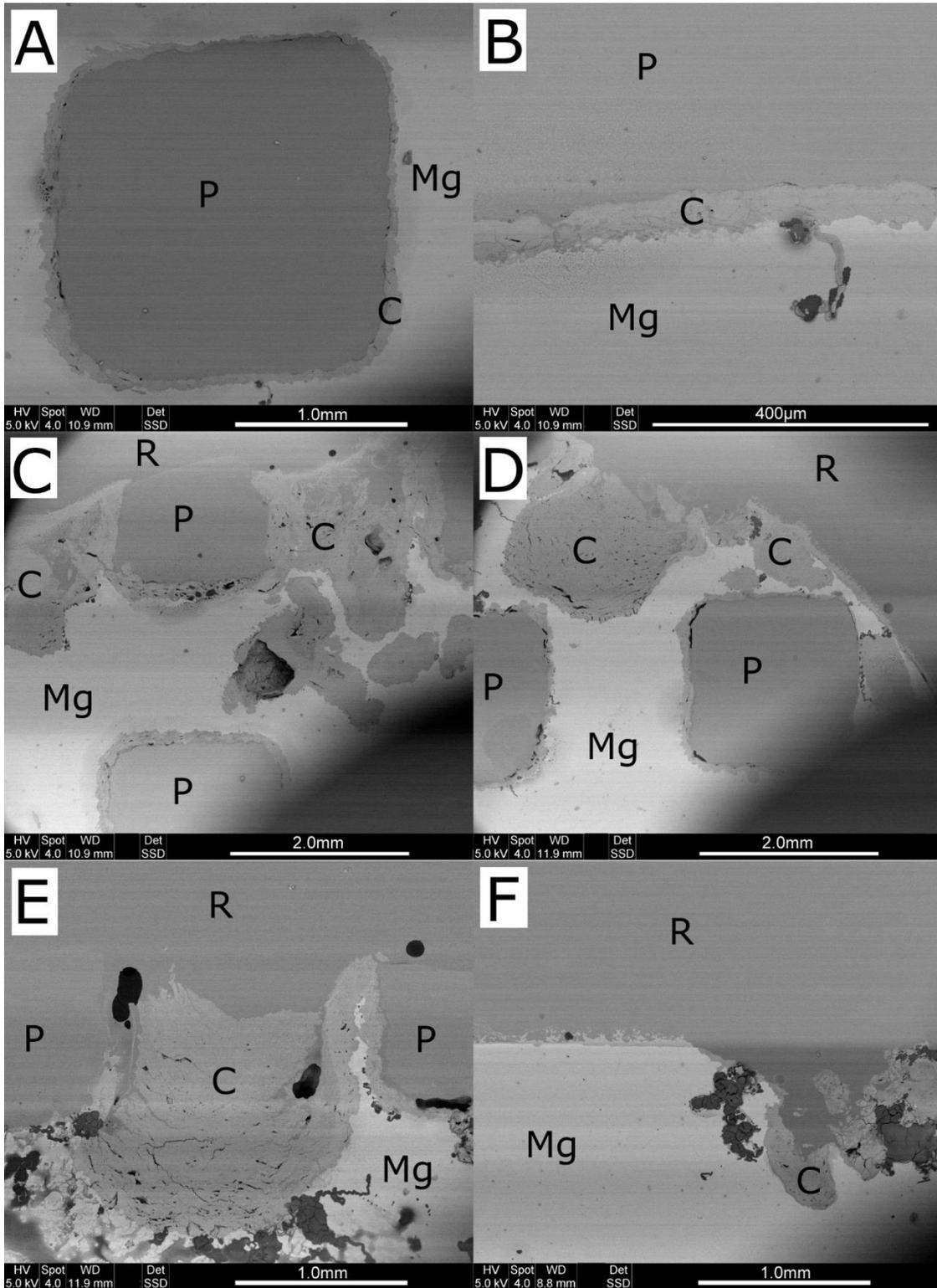

Figure 5: Cross-sectional SEM images (from 5-10 mm into the sample) of the corroded Mg-PLA composite (a-e) and bulk Mg (f) after the immersion test. P = PLA, Mg = Mg, C = corrosion layer, R = mounting resin.

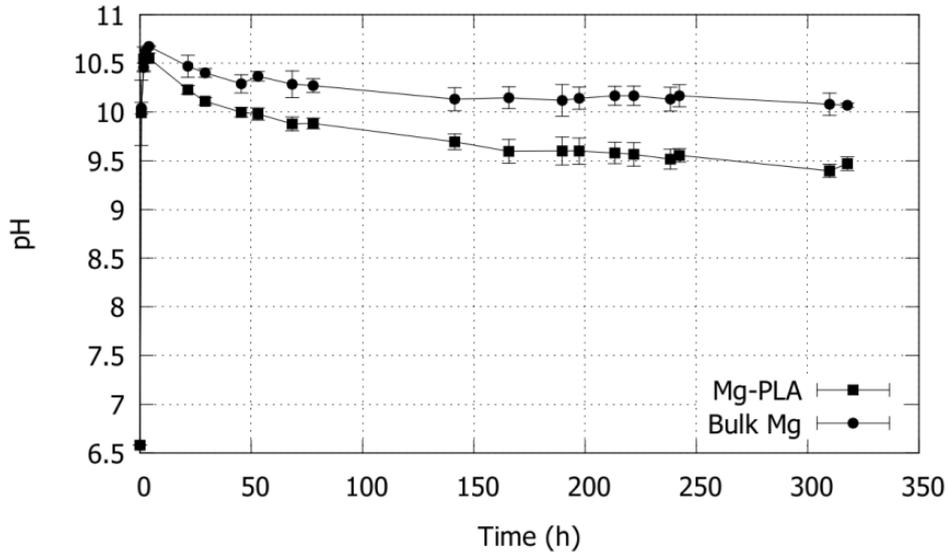

Figure 6: pH trace of the bulk Mg and Mg-PLA composite over the course of the immersion test.

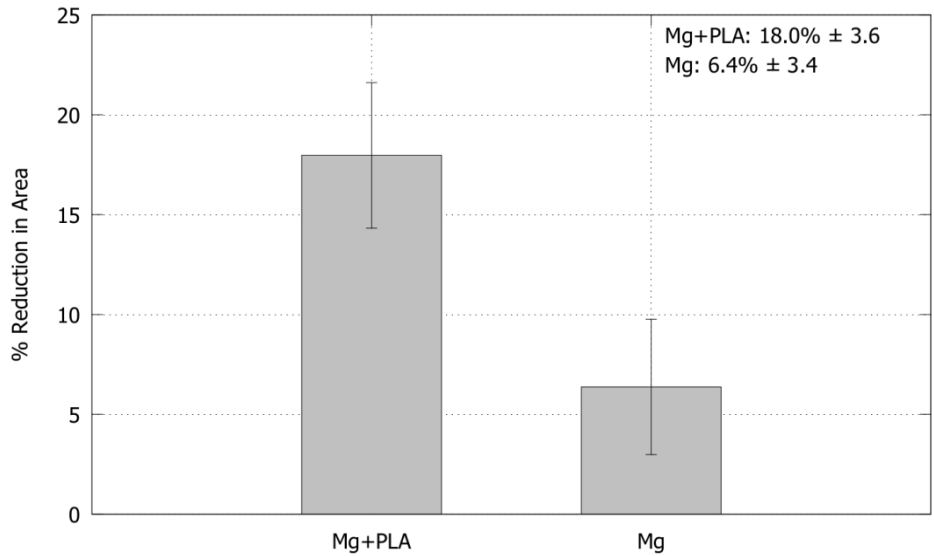

Figure 7: Measured % reduction in area of the bulk Mg and Mg-PLA composite after the immersion test.